\documentclass[a4paper,fleqn]{cas-dc}

\usepackage[numbers]{natbib}

\def\tsc#1{\csdef{#1}{\textsc{\lowercase{#1}}\xspace}}
\tsc{WGM}
\tsc{QE}
\tsc{EP}
\tsc{PMS}
\tsc{BEC}
\tsc{DE}

\begin{document}
\let\WriteBookmarks\relax
\def\floatpagepagefraction{1}
\def\textpagefraction{.001}
\shorttitle{Charming synergies}
\shortauthors{Guy Wilkinson}

\title [mode = title]{Charming synergies: the role of charm-threshold studies in the search for physics beyond the Standard Model}

\author[1]{Guy Wilkinson}[orcid=0000-0001-5255-0619]
\ead{guy.wilkinson@physics.ox.ac.uk}
\address[1]{University of Oxford, Denys Wilkinson Building, Keble Road, OX1 3RH, United Kingdom}

\begin{abstract}
Measurements performed with pairs of charm mesons produced at threshold from the decay of the $\psi(3770)$ resonance are of great value in flavour physics.  The quantum correlation that exists between the two mesons allows unique access to strong-phase information,  which is essential input to flavour-physics studies conducted in other environments. An excellent example from the BESIII collaboration is a recent determination of the strong-phase difference between $D^0$ and $\bar{D}^0$ mesons in the decay $D^0 \to K^0_S\pi^+\pi^-$, which has enabled recent measurements to be performed of the $C\!P$-violating phase $\gamma$ and  $D^0-\bar{D}^0$ oscillations by the LHCb experiment at CERN.
These $\psi(3770)$ data, and also those collected just above the thresholds for $D_s^+$ and $\Lambda_c^+$ production,  can also be exploited in many other ways that are of benefit to flavour-physics studies.  These synergies are reviewed, and the need for larger threshold data samples in the near future is emphasised. 
\end{abstract}


\maketitle

The flavour sector of the Standard Model (SM) of particle physics contains many of the open questions and free parameters of the theory.  Why are there six quarks, with such a wide span of masses?  What determines the hierarchical structure of the Cabibbo-Kobayashi-Maskawa (CKM) matrix? What other sources of $C\!P$ violation exist in nature in addition to that found in the SM, which is accommodated by a complex phase in the CKM matrix?  In recent years, experimental flavour physics has been dominated by studies of beauty and charm hadrons at the LHCb experiment at CERN, with a growing contribution expected in these topics from  the Belle II experiment at KEK, Japan.  However, precise measurements just above the threshold for charm-meson production in $e^+e^-$ collisions, which are possible at the BESIII experiment on the BEPCII accelerator in Beijing, have a very important and complementary role to play in this programme. Recent results from BESIII, mapping out the strong-phase characteristics of certain charm-meson decays, make clear the strong synergies that exist between these diverse facilities.  In particular, precise studies by BESIII of the strong-phase variation in the decay $D^0 \to K^0_S \pi^+\pi^-$ reported in Refs.~\cite{Ablikim:2020yif,Ablikim:2020khq} have been extremely valuable in enabling important measurements of $C\!P$ violation and neutral-meson oscillations at LHCb. 

BEPCII can operate at a collision energy corresponding to the mass of the $\psi(3770)$ resonance, which is the lightest $c\bar{c}$ state that can decay into two charmed mesons.  As no other particles are produced, these mesons exist in a coherent wavefunction. This wavefunction is antisymmetric on account of the quantum numbers of the mother resonance, which are conserved in its decay through the strong interaction.  In the case that the daughter mesons are neutral, they may exist in the flavour eigenstates $D^0$ and $\bar{D}^0$, but there is also the possibility that they are found in a linear superposition of these states, here denoted $D$, of which the $C\!P$-eigenstate case is the simplest example.  The coherence of the wavefunction dictates that if one $D$ meson is reconstructed with even $C\!P$ then the other meson must have odd $C\!P$, and vice versa.  Analogous arguments apply for more general states of superposition.  

This coherent production mechanism allows BESIII to access information in charm-meson decays that is hidden from other experiments.  A hadronic state that can be reached from a $D^0$ decay can also be produced from the decay of a $\bar{D}^0$ meson, with  a difference (in general) in magnitude and phase between the two decay amplitudes.  In the case of a multi-body final state, such as $K^0_S \pi^+\pi^-$ this strong-phase difference (so called, as it is induced by hadronic resonances) will also vary over the kinematics of the decay.  At BESIII direct sensitivity to the phase difference can be obtained by reconstructing both $D$ mesons in the event:  one in the decay of interest, and the other, denoted the `tag' decay, to a $C\!P$ eigenstate  ({\it e.g.} $K^+K^-$ or $\pi^+ \pi^-$). The decay rate contains an interference term that depends on the cosine of the strong-phase difference.  Other categories of tag have sensitivity to the sine of the strong-phase difference.  Following this approach, BESIII has published measurements of the cosine and sine of the strong-phase difference in different regions of phase space for the decay $D^0 \to K^0_S \pi^+\pi^-$~\cite{Ablikim:2020yif,Ablikim:2020khq}.   The analysis is based on the collaboration's current data set of $\psi(3770)$ decays, corresponding to an integrated luminosity of $2.9\,{\rm fb}^{-1}$, which is around four times larger than the sample accumulated around a decade previously by the CLEO-c experiment at Cornell~\cite{Libby:2010nu}. 

\begin{figure}[htp]
    \centering
    \includegraphics[width=0.4\textwidth]{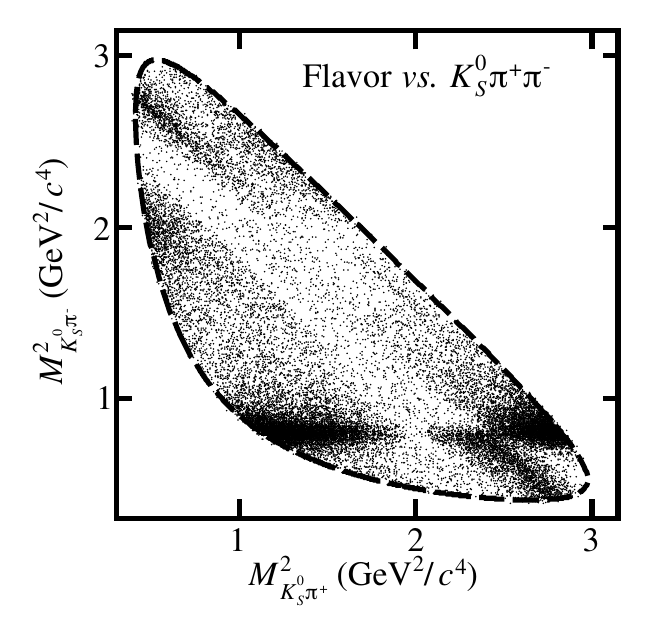}
    \includegraphics[width=0.4\textwidth]{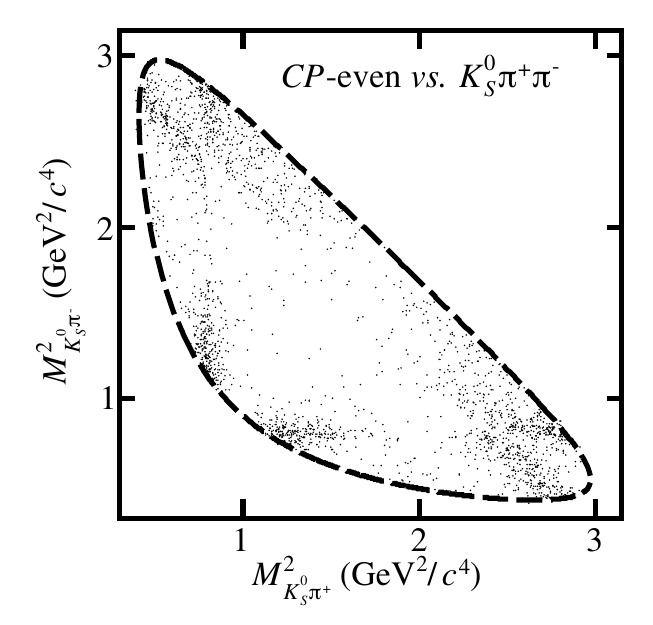}
    \includegraphics[width=0.4\textwidth]{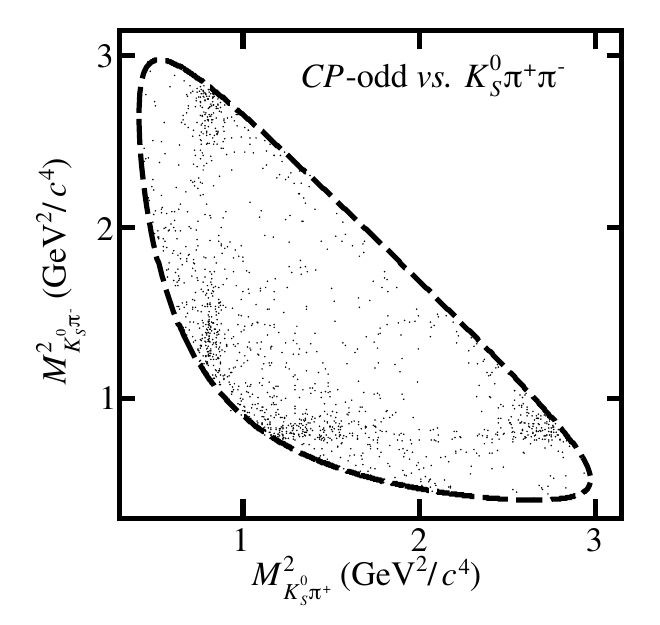}
    \caption{Dalitz plots of the $K^0_S\pi^+\pi^-$ final state from $\psi(3770) \to D \bar{D}$ events at BESIII.  The other meson in the event has been reconstructed in a flavour (top), $C\!P$-even (middle) or $C\!P$-odd (bottom) eigenstate.  Taken from Ref.~\cite{Ablikim:2020khq}.}
    \label{fig:dalitz}
\end{figure}

A visual demonstration of the power of the quantum-correlated charm-pair sample is presented in Fig.~\ref{fig:dalitz}, which displays Dalitz plots for $D \to K^0_S \pi^+\pi^-$ decays in the BESIII analysis. The Dalitz plots show the invariant-mass-squared of the $K^0_S \pi^+$ combination versus that for $K^0_S \pi^-$, with the bands arising from the different intermediate resonances that contribute to the decay.  Each plot corresponds to a separate class of tag, coming from the  accompanying $D$ meson in the event. When the tag decay is to a flavour eigenstate, the charm meson that decays to $K^0_S\pi^+\pi^-$ is also in a flavour eigenstate.  In the other two cases the tag is a $C\!P$ eigenstate, which ensures that the $D \to K^0_S \pi^+\pi^-$ decay is in a state of opposite $C\!P$.  In consequence certain resonances in the plots are enhanced, and others suppressed, with respect to the flavour-eigenstate case.  Analysis of these differences allows the strong-phase characteristics of the decay to be measured.  (Note that the sample size is much larger in the flavour-tag case than for the $C\!P$-eigenstate tags.)

This strong-phase information is of great importance to flavour physicists on account of its practical use in measurements of the angle $\gamma$ of the Unitarity Triangle.  The Unitarity Triangle is the geometrical representation of the CKM matrix in the complex plane, the angles of which can be determined from $C\!P$-violating observables.  Searches for physics beyond the SM can be performed by measuring parameters of the triangle in ways that are expected to be dominated by SM amplitudes, {\it e.g.} through tree-level decays, and comparing with other determinations of the same parameters that are dependent on loop-level processes, and therefore susceptible to the presence of new particles that may modify the values.   The angle $\gamma$ is an excellent example of such a parameter.  At tree-level its value may be determined through studies of interference effects between the decays $B^- \to D^0 K^-$ and $B^+ \to \bar{D}^0 K^-$,  where the $D^0$ and $\bar{D}^0$ mesons decay to a final state common to both.  Comparing the event rates, or other $C\!P$-violating observables, between the $B^-$ and $B^+$ decays allows $\gamma$ to be measured,  but only if the strong-phase characteristics of the charm decay are known.  Last year the LHCb collaboration reported the world's most precise determination of this parameter, exploiting the decay $D \to K^0_S \pi^+\pi^-$ and relying on the BESIII measurement for the strong-phase information~\cite{Aaij:2020xuj}.  The result, $\gamma = \left(68.7^{+5.2}_{-5.1}\right)^\circ$ is in good agreement with the indirect determination of $\gamma = \left(65.7^{+0.9}_{-2.7}\right)^\circ$ coming from all other constraints in the CKM matrix~\cite{Charles:2004jd}. Further improvement is desirable and will come when the LHCb data sample grows, such that the precision of the measurement matches better that of the indirect determination, but the importance of the BESIII results in allowing this study to be performed cannot be overstated.

The same BESIII strong-phase results have been critical inputs to a second noteworthy flavour-physics analysis of LHCb. Very recently the collaboration reported an extremely precise study of $D^0 - \bar{D}^0$ oscillations, which has led to the world's first observation of a non-zero mass difference between the neutral charm eigenstates~\cite{Aaij:2021aam}.  This development is important in paving the way for future searches for $C\!P$ violation associated with charm oscillations, which if seen at a sufficiently high level would signal physics beyond the SM.  The strong-phase information from BESIII allows the analysis to be performed in a model-independent manner.  If these results were unavailable, the LHCb measurement would have been significantly limited in its sensitivity.

Other measurements of strong-phase differences, and associated quantities, can be performed with the same $\psi(3770)$ data set.  A recent study of the decays $D^0 \to K^\mp \pi^\pm \pi^+\pi^-$ has yielded results that are expected to have a similar impact on $C\!P$-violation and oscillation studies at LHCb and Belle II to the analysis already discussed~\cite{Ablikim:2021cqw}, and strong-phase measurements have also been reported for the decays $D^0 \to K^0_SK^+K^-$~\cite{BESIII:2020hpo} and $D^0 \to K^\mp\pi^\pm$~\cite{BESIII:2014rtm}.  In addition to quantum-correlated studies, the very clean environment of $\psi(3770)$ decays allows other interesting topics to be investigated. This statement is equally true of studies made with $D_s^+$ mesons and $\Lambda_c^+$ baryons, at and close to threshold. For these, BESIII has collected samples of $0.5\,{\rm fb}^{-1}$, $3.2\,{\rm fb}^{-1}$ and $0.6\,{\rm fb}^{-1}$ at $\sqrt{s} = $ 4.009~GeV, 4.178~GeV and 4.600~GeV, respectively.  A restricted list of important topics includes:
\begin{itemize}
    \item Absolute and precise measurements of hadronic branching ratios, particularly of `standard candle' decays, such as $D^0 \to K^-\pi^+$~\cite{BESIII:2018apz} or $\Lambda^+_c \to pK^-\pi^+$~\cite{BESIII:2015bjk}, which are valuable for limiting systematic uncertainties in a wide range of flavour-physics analyses.
    \item Studies of the leptonic decays, $D^+_{(s)} \to l^+ \nu_l$ ($l$=$\mu$ or $\tau$).
    Here the measured branching fraction can be used either to determine the decay constant $f_{D^+_{(s)}}$, which then can be compared to the predictions of lattice QCD, or to determine the magnitude of the CKM element $V_{cd(s)}$, or to test lepton-flavour universality through comparing the muonic and tauonic decay rates (see Ref.~\cite{BESIII:2021bdp} for a recent example, concerning $D_s^+ \to \tau^+\nu_\tau$).
    \item Studies of semi-leptonic decays, which are of interest for similar reasons to the leptonic decays, but here with a form factor parameterising the hadronic aspects of the transition (see Ref.~\cite{BESIII:2021duu}  for a recent example, concerning inclusive $D_s^+ \to X e^+ \nu_e $ decays). 
\end{itemize}
Further information on these measurements, and others, can be found in Ref.~\cite{Li:2021iwf}.

Data taking at the $\psi(3770)$ resonance has importance for flavour physics in the lepton sector also.  Radiative returns allow the cross section of processes such as $e^+e^- \to \pi^+\pi^-$ to be measured in the energy region of a few 100\,MeV, which is very important for constraining the hadronic-vacuum-polarisation contribution in the SM calculation of $(g-2)_\mu$, the anomalous magnetic moment of the muon. This observable has received renewed attention in the light of a recent intriguing experimental update from Fermilab~\cite{Muong2:2021ojo}.  BESIII measurements have already provided very important input to these studies~\cite{Ablikim_2021}.  Studies of hadronic states produced in gamma-gamma interactions at these energies are also of interest as they may be used to learn about the hadronic light-by-light contribution. 
Given that the measurement of $(g-2)_\mu$ is expected to improve significantly in the coming years it is vital that the precision of the SM calculation improves also, which would be facilitated through future BESIII measurements based on a data set around four-to-five times larger than is currently available.

Larger samples of threshold data are required not only for $(g-2)_\mu$, but for almost all the topics so far discussed.
It is estimated that with the results from the existing $\psi(3770)$ data set the associated strong-phase uncertainty on the LHCb $\gamma$ measurement will be around $1^\circ$~\cite{Malde:2223391}.  This is less than the current statistical uncertainty, but is expected to become dominant as more $pp$ collision data are accumulated over the coming decade.  A similar story will apply with Belle II, and also for charm-oscillation measurements at both facilities. However, a five-fold increase in the number of recorded $\psi(3770)$ decays, corresponding to a total integrated luminosity of around $20\,{\rm fb}^{-1}$, would enable this improvement in precision from the beauty data to be matched by the charm-threshold data set.
This larger sample, and also those of $D_s^+$ mesons and $\Lambda_c^+$ baryons just above threshold, would also be very valuable for other studies in hadronic flavour physics.  
A more extended discussion on the physics benefits arising from future running at these energies may be found in Ref.~\cite{BESIII:2020nme}.  Collecting these additional decays should be seen as a priority for BESIII, and would enable the experiment to consolidate its unique role in the quest for signs of new physics phenomena through precise measurements in the flavour sector.  Looking to the further future, exciting prospects exist for the construction of one or more `super tau-charm factories' that would allow for even larger data sets to be accumulated, leading to a corresponding improvement in sensitivity across this rich programme of physics~\cite{Luo:2019xqt,EIDELMAN2015238}. 

\vspace{0.3cm}
\noindent {\bf Acknowledgements}
\vspace{0.1cm}

\noindent I am grateful to Achim Denig, Jim Libby and Sneha Malde for stimulating discussions.

\printcredits

\bibliographystyle{JHEP}

\bibliography{cas-refs}





\end{document}